\documentclass[twocolumn]{article}

\usepackage{palatino}
\usepackage{a4}
\usepackage{tabularx}
\usepackage{graphicx}
\usepackage[round]{natbib}
\usepackage{eulervm}
\usepackage{amssymb,amsmath}
\usepackage{url}

\renewcommand{\normalsize}{\fontsize{9pt}{12pt} \selectfont}

\title{Wikipedia: organisation from a bottom-up approach}
\author{Sander Spek \and Eric Postma \\ \\ \normalsize MICC-IKAT, Universiteit
Maastricht\\ \normalsize s.spek@micc.unimaas.nl \and H. Jaap van den Herik}
\date{Paper for the workshop \emph{Research in Wikipedia}, on
the Wikisym 2006.}

\begin{document}

\maketitle

\begin{quotation} \footnotesize \textbf{Abstract.} \it Wikipedia can be
considered as an extreme form of a self-managing team, as a means of
labour division.  One could expect that this bottom-up approach, with
the absense of top-down organisational control, would lead to a chaos,
but our analysis shows that this is not the case.  In the Dutch
Wikipedia, an integrated and coherent data structure is created, while
at the same time users succeed in distributing roles by self-selection.
Some users focus on an area of expertise, while others edit over the
whole encyclopedic range. This constitutes our conclusion that
Wikipedia, in general, is a successful example of a self-managing team. 

\end{quotation}

\section{Work organisation}

For decades, the division of labour has been an essential concept for
people wishing to collaborate in an organisation. This has already been
noted by \nocite{PlatoRepublic}Plato (approx. 390 \textsc{bc}): ``And if
so, we must infer that all things are produced more plentifully and
easily and of a better quality when one man does one thing which is
natural to him and does it at the right time, and leaves other things.''
\citet{SmithWealthNations} attributes great value to the division of
labour too: ``The greatest improvements in the productive powers of
labour, and the greater part of the skill, dexterity, and judgment, with
which it is anywhere directed, or applied, seem to have been the effects
of the division of labour.'' Obviously, this calls for collaboration.
However, according to \citet{Mintzberg1999}, there is a catch: the
division of labour also requires a coordination of labour. The
traditional way to coordinate was by means of a superior, who had either
to simply divide labour and to monitor it, or to manage a team of
people. In the literature from the past decennia, an alternative to this
tradition has arisen: self-managing teams. The Wikipedia community can
perhaps be seen as an ultimate kind of self-management.

Self-managing teams are also called autonomous task groups,
self-managing groups, or empowered groups. They are subgroups of an
organisation, and have been given a high level of autonomy to perform a
full range of tasks. They are expected to ``improve the competence of an
organization to deal with changing environmental demands''
\citep{Balkema1999}. \citet{Daft1998} gives a more extended description
of their expected use. The main improvements are in speed and
efficiency, resulting in a better customer satisfaction. In Wikipedia,
new developments are added uncomparably fast when related to other
encyclopedias. To a reader, this gives Wikipedia an advantage over the
other encyclopedias. Daft also mentions more communication and
cooperation between divisions, increase in enthousiasm of employees
\---which is crucial for a project in which the participants work on a
voluntary base, like in Wikipedia\---, and a decrease of managerial
overhead. Daft has two objections when considering self-managing teams.
The first one is the need for radical changes in the organisation's
structure when making the transition to self-managing teams. However,
Wikipedia never worked in a `traditional way', so a transition is not an
issue. A second objection is the notion that the abilities of managers
and employees to work in these kinds of situations are crucial. Not all
managers and employees might be capable to cope with it. However,
Wikipedia hardly has any managers, and the employees are subject to a
self-selecting mechanism: people that cannot work in `the wiki way' will
drop out by themselves sooner or later, or will maybe not even join.
Therefore, we might expect the Wikipedia `employees' to be well able to
work in a self-managing team.

\section{Organised content}

One might expect that an `unorganised team', like the Wikipedia
community, will produce output that is incoherent and that the work of
some will not fit to the work of others. To test this hypothesis for
Wikipedia, we have studied the article collection of the Dutch
Wikipedia. We can consider this collection to be a network, in which the
articles are nodes and the links between articles are the vertices
between them. This allows us to compare the Wikipedia article network to
other types of networks.

\subsection*{Degrees}

The links in the network of Wikipedia articles are directed.  When there
is a link from A to B, that does not necessarily mean there is a link
from B to A.  For each article we can calculate the the number of
ingoing links (indegree) and the number of outgoing links (outdegree).
The sum of the indegree and outdegree is the degree, a measurement for
the connectedness of a network node. For the nodes in the Dutch
Wikipedia, in June 2005, the average degree was 20.3.  We see that there are many
articles with a low degree and few with a high degree. The distribution
of degrees follows a power law, which is confirmed by \citet{Zlatic2006}.

Authority nodes are articles with an exceptionally high degree. We can
identify several types of authority nodes. When we create a list of the
most referred-to and the most referring articles\footnote{For all experiments, we only
consider the main namespace.  This means that links to and links from
talk pages, special pages, and other administrative pages have not been
taken into account}, we can see a pattern: articles that refer to many
other articles are mostly lists (27 times in the top 50), A to Z pages
(7 times), years or months (7 times), or other overview articles, such
as \emph{Phenomenology of religion}\footnote{We have translated article
names into English.} (which is a small introduction text and a list of
links) and \emph{National anthem} (which at the time included links to
the national anthems of all countries). On the other hand, articles that
are referred to frequently are time units (years and centuries, 10 times
in the top-50), geographical entities (countries, cities, continents: 13
times), and items that have links in templates (such as \emph{biological
kingdom} and \emph{class}, or \emph{zip code} and \emph{e-mail address}:
22 times). Some of the few exceptions in the top-50 are \emph{Second
World War} and \emph{Sport}.

Using the degree, we can divide the nodes, the Wikipedia articles, into
four categories, as indicated in table \ref{table:authorityterminology}.

\begin{table*}
\begin{center}
\begin{tabular}{| c || c | c |} \hline
 & \footnotesize {\bf high indegree} & \footnotesize {\bf low indegree} \\\hline\hline
 \footnotesize {\bf high outdegree} &  \footnotesize all-round authority
& \footnotesize referring authority   \\ \hline
 \footnotesize {\bf low outdegree} &   \footnotesize guru authority & \footnotesize regular node \\ \hline
\end{tabular}
\caption{Terminology for distinguishing articles, based on indegrees and outdegrees.}
\label{table:authorityterminology}
\end{center}
\end{table*}

\begin{enumerate}

   \item \emph{All-round authorities} are articles with both a high indegree and
   a high outdegree. They get referred to frequently, and on their turn, also refer
   readers to other articles.
   
   \item \emph{Guru authorities} are articles with a low outdegree and a high
   indegree. They will probably provide valuable content, as so many articles
   link to them. Examples are \emph{visual arts},
   \emph{universe} and \emph{biological virus}. They describe well-known
   concepts, but do not refer to many other articles. Also, they are the
   articles that get referred to frequently in templates.
   
   \item \emph{Referring authorities} are articles with a high outdegree
and a low indegree. These articles are not referred to frequently, but they
contain many links to other articles. They might provide a good starting
point for readers who look for more specialistic information on a topic.

   \item \emph{Regular nodes} are articles with a low indegree and a low
   outdegree. They constitute the large collection of (semi-)specialised
   articles.

\end{enumerate}

\begin{table*}
\begin{center}
\begin{tabular}{| c || c | c |} \hline
 & {\bf high indegree} & {\bf low indegree} \\\hline\hline
 {\bf high outdegree} & 5,442 articles & 3,834 articles \\ \hline
 {\bf low outdegree} & 3,800 articles & 79,837 articles \\ \hline
\end{tabular}
\caption{Classification of Wikipedia articles in low indegree or high indegree, and low outdegree or high outdegree. A degree is considered high when it is higher than 90\% of the degrees.}
\label{table:inoutdegreecats}
\end{center}
\end{table*}

\subsection*{Clustering and small-worldliness}

\begin{figure}
\begin{center}
\includegraphics[width=0.40\textwidth]{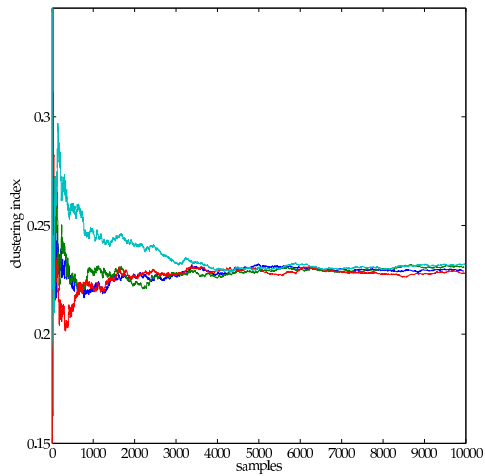}
\caption{The output of four runs of sampling the clustering index.}
\label{fig:clustind4runs}
\end{center}
\end{figure}

An interesting network feature is the clustering index. We found that
the network of the Dutch Wikipedia is too big to calculate the complete
clustering index.  Therefore, we have taken samples by calculating the
clustering index of a randomly selected node. After 50.000 nodes, the
average clustering index seems to stabilise. The output of four runs is
displayed in figure \ref{fig:clustind4runs}. From this data, we conclude
that the clustering index of the Dutch Wikipedia is 0.23, indicating a
fair amount of clustering. This indicates the presence of expertise
fields in the Wikipedia content network.

A high clustering index is one of the two characteristics of small-world
networks. \citep{Watts1998}. The other feature is the average shortest
path between two random notes. When we have calculated this, and thereby
concluded that the Wikipedia network is a small-world network, this
would bring interesting conclusions. A small-world network has several
benefits, as discussed by \citet{Kleinberg2000}. For Wikipedia, we see
benefits in short navigation paths, offering browsing as an alternative
to searching to users.

\subsection*{Scale-freeness}

Scale-free networks are networks with a power-law degree distribution
\citep{Barabasi1999,Newman2003}, which means that the number of nodes having
$n$ links decreases exponentionally, starting from $n=1$. In a formula, this is
denoted by $P(v_{n}) \propto n^{-\alpha}$, where $P$ is the probability of a
vertex $v$ having a degree of $n$. This type of network is characterised by a
small number of highly connected nodes (thus having a high degree), whereas
most nodes have a low degree.  The high-degree nodes act as connection points
between the different nodes of the network.  The exponent of the network,
$\alpha$, can be seen as a measurement for the scale-freeness of a network.
Most scale-free networks have an $\alpha$ between two and three. Networks that
conform to this $\alpha$ are amongst others citation networks, the Internet,
and the World Wide Web \citet[page 10]{Newman2003}.

\citet{Barabasi1999} explain the phenomenon of scale-freeness by two generic
mechanisms: (1) the network typically expands by the addition of new vertices,
and (2) new vertices tend to connect to high-degree vertices. For Wikipedia,
these mechanisms apply, since the addition of new articles shows a steady
growth\footnote{\url{http://en.wikipedia.org/wiki/Wikipedia:\\Modelling_Wikipedia's_growth}},
and new articles generally link to well-connected vertices such as countries and
years.

\begin{figure}
\begin{center}
\includegraphics[width=0.45\textwidth]{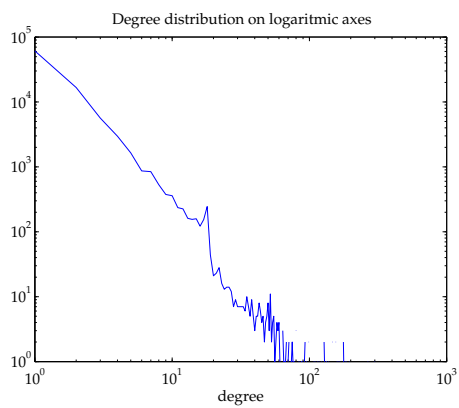}
\caption{Logaritmic plot of the degrees of all the vertices of the Dutch Wikipedia. The plot fits the function $y=2.1 \cdot 10^{5}e^{-1.24x}$.}
\label{fig:degreeplot}
\end{center}
\end{figure}

A plot with logaritmic scales of the degrees of all the vertices in the network
is displayed in figure \ref{fig:degreeplot}. In the figure, we can see that the
number of nodes having $n$ links decreases exponentially. The function that can
be fitted to this distribution is $2.1 \cdot 10^{5}e^{-1.24x}$. This means the
scale-free-network exponent is of value 1.24. Compared to the other networks
mentioned by \citet[page 10]{Newman2003}, we can see that Wikipedia has the
lowest scale-free exponent of all the networks. This means that Wikipedia has
the characteristics of scale-freeness, but in a less radical way then the other
networks.

\section{Organised work division}

In the real world, authors develop an expertise. They study a specific
area of knowledge and during the process they are seen more and more as
an authority in this field by others. Also in Wikipedia, many users
restrict themselves to a certain area of expertise. Therefore, we will
study to see how the expertise of authors maps onto Wikipedia domains.
We will do this by identifying certain expertise fields in the Wikipedia
knowledge collections, and consequently see if author's contributions
are evenly scattered among these fields, or if they rather cluster in
fields of expertise.

Wikipedia articles are tagged by categories that indicate the nature of the subject. An extensive analysis of Wikipedia classes can be found in work by \citet{Voss2006}. We have manually selected fourty categories as a broad mixture of categories that
can be found in Wikipedia. The subjects range from science and
social/historical topics to culture and sports. Some categories are general
(e.g., physics), while others are more specialised (e.g., Spanish chess
player). Some categories refer to the Dutch-speaking area (e.g., Belgian
political party), while others are about more `exotic' regions (e.g., Mexico).

The fourty categories are grouped into five classes, namely science,
social/historical, culture, geography, and sports.

\subsection*{Expertises in categories}

In order to quantify the differences between the categories, we have
taken two statistical measurements: (1) the number of edits and the
number of unique authors, resulting in the average edits per author
($\overline{ea}$), and (2) a Pareto analysis. The formula for
$\overline{ea}$ is (adapted from \citet{McClave1998}):

\begin{equation*}
   \overline{ea} = {\displaystyle\sum_{i=1}^n{{ea}_i} \over n}
\end{equation*}

As described by amongst others \citet[p. 31]{McClave1998} and
\citet{Reed2003}, Pareto-analysis checks for the so-called
Pareto-principle\index{Pareto!principle}: a power-law
distribution\index{distribution!power law} where the larger part of the
consequences is generated by a small part of the causes. This is also
called ``the vital few, and the trivial many'', or in more popular
terms, the eighty-twenty rule\index{eighty-twenty rule}. The Italian
economist Vilfredo Pareto\index{Pareto!Vilfredo} (1843\--1923)
discovered this rule when he found that approximately 80 per cent of the
wealth of a country lies with approximately 20 per cent of the
population. According to \citet[p.  31]{McClave1998}, V. E. Kane found
similar patterns for other (economic) areas, such as 80 per cent of
sales being attributable to 20 per cent of the customers, or 80 per cent
of the customer complaints referring to 20 per cent of the components.
The Pareto
distribution\index{distribution!Pareto}\index{Pareto!distribution} is
comparable to other power laws, such as Zipf
distributions\index{distribution!Zipf} \citep{Newman2000,Reed2001}. To
perform a Pareto analysis, we will gather the relative number of edits
(consequences) resulting from the top 20\% of the editors (causes).

The number of edits per category range from 370 edits (Belgian political
party), to 15,396 edits (mathematics). The number of different authors
ranges from 13 (Spanish chess player) to 280 (physics). As a result,
$\overline{ea}$ lies between 5.8 (Belgian political party) and 382.8
(Spanish chess player). In the latter case, 4977 edits have been made,
by only those 13 authors we just mentioned. In general, we can say the
articles with a high $\overline{ea}$ are the more specialistic articles,
with topics most people will not be able to tell much about. Except for
the two mentioned topics, this also includes chess player (272.1),
translator (256.3), Russian political party (214.0), and peace treaty
(147.0). The articles with a low $\overline{ea}$ deal with topic areas
that most people have at least some expertise in, or topic areas that
everyone claims to know about. This includes amongst others investing
(7.5), cartography (10.3), cult movie (10.6), and philosophy (11.3).
Cartography seems to be the only exception to the pattern described. The
average $\overline{ea}$ is 92.5

When we look for a Pareto-principle, we find that on average the top-20
authors account for 67\% of the edits. Low scores are for chess player
(21.3), Spanish chess player (42.7), and French chess player (46.2).
Highest scorers are physics (82.7), literature (78.3), and politics
(77.5). Based on this data, one could claim that the chess categories
are therefore not really specialistic, since there is no `elite' that
accounts for most of the edits. However, when we look at the
contribution of the top author, the top-1, we see that at least the
Spanish chess players have one major contributor, who did 29\% of all
edits in that category. Hence, we might argue that the number of edits
per author in this category declines even more exponential than in a
Pareto curve. The expertise lies with less than 20\% of the editors in
the category. Other categories that have \emph{gurus}\index{guru}, users
that account for a high percentage of the edits, are philosophy (29.6\%)
and Russian political party (38.1\%). On average, the most active user
per category is responsible for 17.4\% of the edits.

\subsection*{Expertises of authors}

In the previous section, we started our analysis from the viewpoint of
the category, and studied the distribution of edits over the authors. In
this section, we will start from the author's point of view, and study
his distribution of edits over the categories. We took the same 40
categories as described in the previous section, and took into account
any user that has made at least one edit in any of the categories.

\begin{figure}
\begin{center}
\includegraphics[width=0.45\textwidth]{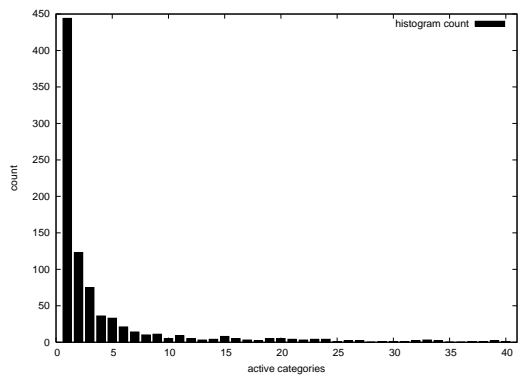}
\caption{Histogram of the number of active categories for users.}
\label{fig:autcathisto}
\end{center}
\end{figure}

First, we studied in how many categories users typically are active. Our
definition of `active' is very weak: we count a user as being active in
a category when the users has made at least one edit in that category.
The most users, 444 out of 856, are active in only one category. Only
one user seems to be active in all categories, but this is the user who
has user id 0 in the database. This is the cumultative of all anonymous
users. Still, there are two users active in all-but-one category. The
total histogram of the number of active category per user follows a power law, as is displayed
in figure \ref{fig:autcathisto}.

\begin{figure}
\begin{center}
\includegraphics[width=0.45\textwidth]{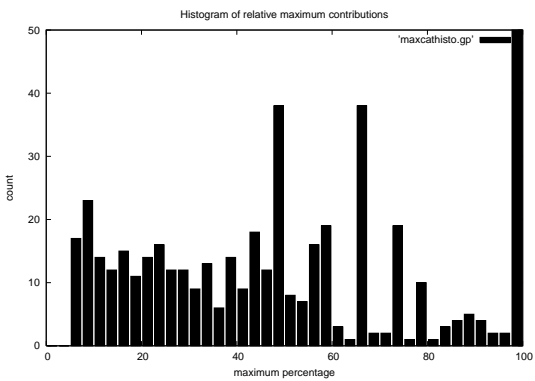}
\caption{Histogram of the relative maximum contributions.}
\label{fig:maxcathisto}
\end{center}
\end{figure}

When we consider expertise, we might take a look at the category that authors
make their most edits in. We have calculated the contribution of each author in
its most active category, relative to the author's total contributions in all
the used categories. Of course, all the authors that have made only one edit,
now score a 100\% maximum percentage. Overall, the maximum percentages are
distributed as in figure \ref{fig:maxcathisto}. There is no clear pattern,
although less authors seem to have a high maximum percentage, apart from the
one-edit authors.

Another measurement for the distribution of an author's edits over categories is
\emph{entropy}, an application of the concept of \emph{information entropy} as
invented by \citet{Shannon1951}. For each author, we have calcultated the number
of edits in a certain category relative to the total number of edits of that
author ($p_{a,c}$). We calculated the entropy of an author ($H_a$) as follows:

\begin{equation*}
   H_a =  {\displaystyle\sum_{c} p_{a,c} log_2 (p_{a,c})}
\end{equation*}

\begin{figure}
\begin{center}
\includegraphics[width=0.45\textwidth]{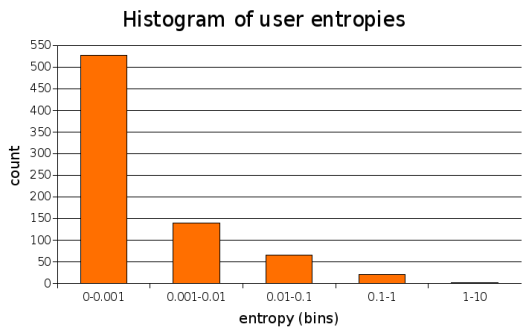}
\caption{Histogram of the author entropies.}
\label{fig:authorentropyhisto}
\end{center}
\end{figure}

In this way, we end up with a list of the author entropies of all 856
authors from the sample, ranging from 0.00005 to 5.0075. The entropy of
the collective of anonymous users equals the maximum of 5.0075. The
average entropy is 0.0182. A histogram of all entropies is displayed in
figure \ref{fig:authorentropyhisto}. In this figure, we see that most of
the users have a low entropy, but there also exist users with higher
entropies. This confirms our belief that there are two types of users:
those who edit in a certain field of expertise, and those who edit
througout the whole Wikipedia. The users in the last category will
mostly be the users with much general knowledge or the users who perform
administrator tasks.

\section{Conclusions and discussion}

In this paper, we have studied Wikipedia as a self-managing team. It
lacks top-down control, which could lead to chaotic output and bad
coordination. Our analysis of the Dutch Wikipedia shows that this is
not the case. The network of Wikipedia articles shows clustering,
scale-freeness, and perhaps even small-worldliness. Articles with
a high number of ingoing or outgoing links are crucial in this network.

When studying the distribution of edits over the authors, we can
distinguish categories of articles that are more or less specialistic.
We can also make the same distinction on authors by using the entropy of
the distribution of their edits over the categories. We find that some
authors only edit in typical specialistic categories, while other
authors edit over the whole range of articles. The latter are presumably
people with more general knowledge or administrators who check for
vandalism and obvious errors.

The data in this paper provides in interesting starting point for more
research on article types and author types, and especially the mapping
between the two.

\subsection*{Acknowledgements}

This work has been performed with support and advise of Antal van den Bosch and Jakob Voss.

\bibliographystyle{natbib}
\bibliography{literature}

\end{document}